\documentclass{article}
\usepackage[left=2.5cm, top=3cm, bottom=3cm, right=2.5cm]{geometry}
\usepackage{amsfonts}
\usepackage{amsthm}
\usepackage{amsmath}
\usepackage{enumerate}
\usepackage[numbers,sort&compress]{natbib}
\usepackage[dvips]{graphicx}

\newcommand{\dd}{\mathrm{d}}

\newcommand{\ud}{\,\dd}
\newcommand{\nbr}[1]{$#1$\nobreakdash-\hspace{0pt}}
\newcommand{\braket}[1]{\langle{#1}\rangle}

\newcommand{\lshad}{[\![}
\newcommand{\rshad}{]\!]}
\newcommand{\sdot}{\,\cdot\,}
\providecommand{\abs}[1]{\lvert#1\rvert}

\numberwithin{equation}{section}

\title{Coherence and squeezing along quantum trajectories}
\author{Ziemowit Doma\'nski\\
\small Center for Theoretical Physics of the Polish Academy of Sciences\\
\small Al. Lotnik\'ow 32/46, 02-668 Warsaw, Poland\\
\small \tt domanski@cft.edu.pl \and Maciej B{\l}aszak\\
\small Faculty of Physics, Division of Mathematical Physics, A. Mickiewicz
University\\
\small Umultowska 85, 61-614 Pozna\'n, Poland\\
\small \tt blaszakm@amu.edu.pl}

\begin{document}

\maketitle

\begin{abstract}
We perform a detailed analysis of the behavior of coherent and squeezed states
undergoing time evolution. We calculate time dependence of expectation values
of position and momentum in coherent and squeezed states (which can be
interpreted as quantum trajectories in coherent and squeezed states) and examine
how coherence and squeezing is affected during time development, calculating
time dependence of position and momentum uncertainty. We focus our
investigations on two quantum systems. First we consider quantum linear system
with Hamiltonian quadratic in $q$ and $p$ variables. As the second system we
consider the simplest quantum nonlinear system with Hamiltonian quartic in $q$
and $p$ variables. We calculate the explicit formulas for the time development
of expectation values and uncertainties of position and momentum in an initial
coherent state.
\\[\baselineskip]
\end{abstract}

\section{Introduction}
\label{sec:1}
Coherent and squeezed states play an important role in quantum optics
\cite{Glauber:1963,Glauber:1964,Sudarshan:1963,Walls:1983,Han:1988}. They are,
moreover, the closest analog of classical states. However, the coherence of
states is a property which is not preserved during time evolution of most
quantum systems. In fact, for a physical quantum system, states undergoing time
development often finally approach one of the stationary states of the system.
For this reason in quantum mechanics one is mostly dealing with stationary
states (eigen-vectors of the Hamilton operator).

In classical Hamiltonian mechanics solutions of Hamilton equations represent
trajectories in phase space, which are important tool in investigation of
geometry of Hamiltonian systems. The analog of classical trajectories can be
also formulated in quantum mechanics \cite{Blaszak:2012b}. In classical case
trajectories represent time development of points in phase space (classical
coherent states). However, in quantum case points in phase space loose physical
interpretation, and thus trajectories itself are not ``physical'' objects.
Nevertheless, quantum trajectories can be used to calculate time dependence of
expectation values of observables which have a direct physical interpretation.
In fact, solutions of quantum Hamilton equations, which can be interpreted as
quantum trajectories (represented in some coordinates), describe time
development of position and momentum observables. Taking expectation values of
position and momentum results in a trajectory in a quantum state, which
describes average path along which a particle in phase space will move.

In this paper we would like to investigate in more details the behavior of
coherent and squeezed states during time development. In particular, we will
calculate time dependence of expectation values and uncertainties of position
and momentum in coherent and squeezed states (quantum trajectories in coherent
and squeezed states) and examine how coherence and squeezing is affected during
time development.

The analysis will be performed for two fairly different quantum systems. The
first system is a linear quantum system, described by a general Hamiltonian
quadratic in $q$ and $p$ variables, so in that case quantum trajectories
coincide with their classical counterpart. This is one of the simplest
Hamiltonians we can consider and has the form of the Hamiltonian of harmonic
oscillator with some interaction term. Such Hamiltonian is very often found in
quantum optics where it describes nonlinear interactions of light with a medium.
In practice, such nonlinear effects are used to create squeezed states of light.
For such Hamiltonian time evolution of expectation values of position and
momenta coincide with quantum trajectories and, as we show in
Section~\ref{sec:3}, even in this case coherence is not always preserved in
time.

The second system, represents nonlinear quantum system, described by a
Hamiltonian quartic in $q$ and $p$ variables. It is an example of a system with
purely quantum time evolution, i.e. the solutions of quantum Hamilton equations
do not coincide with their classical counterpart. This is an interesting example
of pure quantum flow, revealing the property that evolution of such systems is
not well defined for all values of the evolution parameter $t$.

Singularities of classical trajectories are not admissible as each classical
trajectory represents simultaneously expectation values of positions
and momenta of a system in a classical coherent state, i.e. Dirac delta
distribution. Contrary, pure quantum trajectories themselves are not
``physical'' objects as Dirac distributions are not admissible quantum states
so, singularities of pure quantum trajectories are acceptable.

Unfortunately, even when we consider time evolution of expectation values of
position and momentum observables in a quantum coherent state, we get formulas
well defined only on certain intervals of $t$. This is in a strong contrast to
the classical case where such time evolution is defined for all
$t \in \mathbb{R}$. As we show in Section~\ref{sec:4} for this type of systems
the states quickly loose coherence and squeezing during time development.

Throughout the paper we will be using the phase space formalism of quantum
mechanics. In this formalism many results of this paper can be formulated
simpler. It is also the formalism widely used in quantum optics where it is
convenient to consider states of the light as pseudo-probabilistic distribution
functions \cite{Hillery:1984}. A very short description of phase space quantum
mechanics is given in Section~\ref{sec:2}.

\section{Quantum Hamiltonian systems}
\label{sec:2}
A convenient approach to quantum mechanics, which will be useful for
investigating time development of expectation values of position and momentum
observables, is quantum mechanics on phase space \cite{Groenewold:1946,%
Moyal:1949,Bayen:1975-1977,Bayen:1978a,Bayen:1978b,Blaszak:2012}. In
this approach to quantum theory one deforms, with respect to a deformation
parameter $\hbar$ (Planck's constant), a Poisson algebra of a classical
Hamiltonian system to a noncommutative algebra. This new algebra describes then
a quantum Hamiltonian system. One of the benefits of such description of quantum
mechanics is that it uses a similar formalism as classical theory.

In what follows we will focus on systems for which phase space is of the form
$\mathbb{R}^2$. A classical Poisson algebra is then an algebra
$C^\infty(\mathbb{R}^2)$ of smooth complex-valued functions on $\mathbb{R}^2$
with a point-wise multiplication and a standard Poisson bracket. A deformation
of this algebra can be introduced in the following way. The point-wise product
of functions can be replaced by the following noncommutative product
\begin{align}
f \star_M g & = f \exp \left( \frac{1}{2} i\hbar \overleftarrow{\partial}_{q}
    \overrightarrow{\partial}_{p} - \frac{1}{2} i\hbar
    \overleftarrow{\partial}_{p} \overrightarrow{\partial}_{q} \right) g
    \nonumber \\
& = f \cdot g + o(\hbar).
\label{eq:4}
\end{align}
This \nbr{\star_M}product is associative and reduces to the point-wise product
in the limit $\hbar \to 0$. It is usually called the Moyal product
\cite{Moyal:1949}. Moreover, the Poisson bracket $\{\sdot,\sdot\}$ is replaced
by the following Lie bracket
\begin{align}
\lshad f,g \rshad & = \frac{1}{i\hbar} [f,g]
= \frac{1}{i\hbar}(f \star_M g - g \star_M f) \nonumber \\
& = \{f,g\} + o(\hbar).
\end{align}
It indeed reduces to the Poisson bracket $\{\sdot,\sdot\}$ in the limit
$\hbar \to 0$. The Moyal product \eqref{eq:4} can be also written in the
following integral form \cite{Baker:1958}
\begin{equation}
(f \star_M g)(q,p) = \frac{1}{(\pi\hbar)^2} \int_{\mathbb{R}^4} f(q',p')
    g(q'',p'') e^{\frac{2i}{\hbar}((q - q')(p - p'') - (p - p')(q - q''))}
    \ud{q'}\ud{p'}\ud{q''}\ud{p''}.
\label{eq:5}
\end{equation}
The Moyal quantization of a classical Hamiltonian system results in a quantum
theory equivalent to the standard Weyl quantization in the Hilbert space
of states equal $L^2(\mathbb{R})$, being the position representation of the
Moyal quantization.

The algebraic structure of the quantum Poisson algebra fully characterizes
quantum states. It can be shown \cite{Blaszak:2013b} that quantum states can be
represented as pseudo-probabilistic distribution functions, i.e., as square
integrable functions $\rho \in L^2(\mathbb{R}^2)$ defined on the phase space and
satisfying
\begin{enumerate}[(i)]
\item $\rho = \bar{\rho}$ (self-conjugation),
\item $\displaystyle \int_{\mathbb{R}^2} \rho \ud{l} = 1$ (normalization),
\item $\displaystyle \int_{\mathbb{R}^2} \bar{f} \star_M f \star_M \rho \ud{l}
\ge 0$ for $f \in C^\infty_0(\mathbb{R}^2)$ (positive-definiteness),
\end{enumerate}
where $\dd{l} = \frac{\dd{q}\ud{p}}{2\pi\hbar}$ is a normalized Liouville
measure and $C^\infty_0(\mathbb{R}^2)$ is the space of smooth functions on
$\mathbb{R}^2$ with compact support. Pure states can be described as those
functions $\rho_\text{pure} \in L^2(\mathbb{R}^2)$ which are self-conjugated,
normalized and idempotent
\begin{equation}
\rho_\text{pure} \star_M \rho_\text{pure} = \rho_\text{pure}.
\label{eq:21}
\end{equation}
They are in fact the well known Wigner functions \cite{Wigner:1932}.

Real-valued functions in the quantum Poisson algebra play the role of quantum
observables. An expectation value of a quantum observable $A$ in a state $\rho$
is defined in a similar way as in the classical case and is given by the formula
\begin{equation}
\braket{A}_\rho = \int_{\mathbb{R}^2} A \star_M \rho \ud{l}
= \int_{\mathbb{R}^2} A \rho \ud{l}.
\end{equation}
Let us also recall a definition of the uncertainty (standard deviation) of a
given observable $A$ in a state $\rho$:
\begin{equation}
(\Delta A)^2 = \braket{(A - \braket{A}_\rho)^2}_\rho
= \braket{A^2}_\rho - \braket{A}^2_\rho,
\end{equation}
where now squares of observables are taken with respect to the Moyal product,
e.g. $A^2 = A \star_M A$.

Time evolution of a quantum system is governed by a Hamilton function $H$ which
is, similarly as in classical mechanics, some distinguished observable. An
equation of motion of states (Schr\"odinger picture) is the counterpart of the
Liouville's equation describing time evolution of classical states (probability
distributions), and is given by the formula \cite{Blaszak:2012}
\begin{equation}
\frac{\partial\rho}{\partial t}(t) - \lshad H,\rho(t) \rshad = 0.
\end{equation}
Time evolution of a quantum observable $A(q,p)$ (Heisenberg picture) is given by
\begin{equation}
\frac{\dd A}{\dd t}(t) - \lshad A(t),H \rshad = 0.
\end{equation}
In particular for observables of position and momentum $Q(q,p,t=0) = q$,
$P(q,p,t=0) = p$ we get quantum Hamilton equations
\begin{equation}
\frac{\dd Q}{\dd t}(t) - \lshad Q(t),H \rshad = 0, \quad
\frac{\dd P}{\dd t}(t) - \lshad P(t),H \rshad = 0.
\label{eq:6}
\end{equation}
Their solutions
\begin{equation}
\Phi_t(q,p;\hbar) = (Q(q,p,t;\hbar), P(q,p,t;\hbar))
\end{equation}
represent quantum flow $\Phi_t$ in the phase space \cite{Blaszak:2012b}.
For every instance of time $t$ the map $\Phi_t$ is a quantum canonical
transformation (quantum symplectomorphism).

\section{Coherent and squeezed states}
\label{sec:3}
In this section we will characterize the behavior of coherent and squeezed
states under time evolution governed by a general Hamiltonian quadratic in $q$
and $p$ variables. The notion of coherent and squeezed states is often defined
in relation to some system, usually a harmonic oscillator. Thus let us consider
a harmonic oscillator described by a Hamiltonian
\begin{equation}
H(\tilde{q},\tilde{p}) = \frac{1}{2m}\tilde{p}^2
    + \frac{1}{2}m\omega^2 \tilde{q}^2.
\end{equation}
It is convenient to introduce normalized position and momentum observables
\begin{equation}
q = \sqrt{m\omega}\tilde{q}, \quad p = \frac{1}{\sqrt{m\omega}}\tilde{p}.
\end{equation}
In these new variables the Hamiltonian of the harmonic oscillator takes the form
\begin{equation}
H(q,p) = \frac{1}{2}\omega(p^2 + q^2).
\label{eq:9}
\end{equation}
A coherent state of the harmonic oscillator is a state for which uncertainties
$(\Delta q)^2, (\Delta p)^2$ of the normalized position and momentum observables
are equal $\frac{\hbar}{2}$. A state is called squeezed if one of the
uncertainties $(\Delta q)^2$ or $(\Delta p)^2$ is smaller than
$\frac{\hbar}{2}$. Then necessarily the other uncertainty has to be bigger than
$\frac{\hbar}{2}$, so that to satisfy the Heisenberg uncertainty relation.
Coherent states minimize the Heisenberg uncertainty relation, i.e. they satisfy
\begin{equation}
\Delta q \Delta p = \frac{\hbar}{2}.
\label{eq:10}
\end{equation}
Squeezed states do not necessarily have to minimize the Heisenberg uncertainty
relation. If they do they are called ideal squeezed states.

Quantum coherent states have the form of Gaussian functions
\begin{equation}
\rho(q,p) = 2\exp\left(-\frac{(q - q_0)^2}{\hbar}\right)
    \exp\left(-\frac{(p - p_0)^2}{\hbar}\right),
\label{eq:11}
\end{equation}
where $(q_0,p_0)$ is a mean normalized position and momentum in the state
$\rho$. Recall that $\rho$ is normalized with respect to the normalized
Liouville measure $\dd{l} = \frac{\dd{q}\ud{p}}{2\pi\hbar}$. It is also a pure
quantum state as one can check that \eqref{eq:11} fulfills condition
\eqref{eq:21}.

The Hamiltonian of the harmonic oscillator preserves the coherence of states,
i.e. a coherent state remains coherent during time evolution. Indeed, the
quantum Hamilton equations \eqref{eq:6} read
\begin{equation}
\begin{split}
\frac{\dd Q}{\dd t} & = \omega p \frac{\partial Q}{\partial q}
    - \omega q \frac{\partial Q}{\partial p}, \\
\frac{\dd P}{\dd t} & = \omega p \frac{\partial P}{\partial q}
    - \omega q \frac{\partial P}{\partial p}.
\end{split}
\label{eq:23}
\end{equation}
These equations coincide with the classical equations describing time evolution
of the observables of position and momentum $Q(q,p) = q$, $P(q,p) = p$ since
$\lshad H,f \rshad = \{H,f\}$ for every $f$. As a result classical and quantum
trajectories coincide \cite{Blaszak:2012b}. In fact such a situation occurs for
all systems with Hamiltonians being quadratic functions of phase space
coordinates. Thus, to solve \eqref{eq:23} we can first solve the classical
Hamilton equations of the harmonic oscillator
\begin{equation}
\begin{split}
\dot{q} & = \omega p, \\
\dot{p} & = -\omega q,
\end{split}
\end{equation}
which solution describing time development of classical pure states is equal
\begin{equation}
\begin{split}
q(t) & = q_0 \cos\omega t + p_0 \sin\omega t, \\
p(t) & = p_0 \cos\omega t - q_0 \sin\omega t,
\end{split}
\label{eq:22}
\end{equation}
where $(q_0,p_0)$ is the initial position and momentum. By virtue of
\eqref{eq:22} we can deduce that time evolution of $Q$ and $P$, in classical and
quantum case, is expressed by the formulas
\begin{equation}
\begin{split}
Q(q,p,t) & = q \cos\omega t + p \sin\omega t, \\
P(q,p,t) & = p \cos\omega t - q \sin\omega t
\end{split}
\end{equation}
which indeed fulfill equations \eqref{eq:23}. A straightforward calculation
shows that the expectation values of $Q$ and $P$ in a coherent state
\eqref{eq:11} are equal $Q(q_0,p_0,t)$ and $P(q_0,p_0,t)$, which means that
time evolution of expectation values of position and momentum observables in
coherent states coincide with quantum trajectories (in a full analogy with the
classical case). Moreover, the uncertainties $(\Delta Q)^2$ and $(\Delta P)^2$
in a coherent state \eqref{eq:11} are equal $\frac{\hbar}{2}$. Thus, coherent
states remain coherent during time development.

We will investigate how coherent and squeezed states behave during time
evolution given by the Hamiltonian of the harmonic oscillator with the
interaction term of the form
\begin{equation}
H_I(q,p) = \alpha qp + \frac{1}{2}\beta p^2 - \frac{1}{2}\beta q^2,
\end{equation}
where $\alpha,\beta \in \mathbb{R}$ are some constants.
So, let us consider the following Hamiltonian
\begin{equation}
H(q,p) = \frac{1}{2}(\omega + \beta)p^2 + \frac{1}{2}(\omega - \beta)q^2
    + \alpha qp.
\label{eq:24}
\end{equation}
Note, that any Hamiltonian quadratic in $q$ and $p$ variables is of the above
form for some values of constants $\omega$, $\alpha$ and $\beta$. Again the
classical and quantum time evolution equations of observables $Q$ and $P$
coincide, thus the solution of quantum Hamilton equations can be found by
solving the classical Hamilton equations
\begin{equation}
\begin{split}
\dot{q} & = \alpha q + (\beta + \omega)p, \\
\dot{p} & = (\beta - \omega)q - \alpha p.
\end{split}
\end{equation}
To solve this system of differential equations let us write it in a matrix form
\begin{equation}
\frac{\dd}{\dd t} \begin{pmatrix} q \\ p \end{pmatrix} =
    \begin{pmatrix}
    \alpha & \beta + \omega \\
    \beta - \omega & -\alpha
    \end{pmatrix} \begin{pmatrix} q \\ p \end{pmatrix}.
\end{equation}
The solution of the above equation reads
\begin{equation}
\begin{pmatrix} q(t) \\ p(t) \end{pmatrix} = \exp\left[t
    \begin{pmatrix}
    \alpha & \beta + \omega \\
    \beta - \omega & -\alpha
    \end{pmatrix}\right] \begin{pmatrix} q_0 \\ p_0 \end{pmatrix}.
\end{equation}
Direct computation of the exponent in the above equation allows us to find the
solution of the Hamilton equations. Hence, after introducing
$R = \sqrt{\abs{\omega^2 - \alpha^2 - \beta^2}}$ the solution to
the Hamilton equations in a case $\omega^2 > \alpha^2 + \beta^2$ reads
\begin{equation}
\begin{split}
q(t) & = \frac{\alpha q_0 + (\omega + \beta) p_0}{R}\sin(Rt) + q_0\cos(Rt), \\
p(t) & = -\frac{(\omega - \beta) q_0 + \alpha p_0}{R}\sin(Rt) + p_0\cos(Rt).
\end{split}
\end{equation}
When $\omega^2 < \alpha^2 + \beta^2$ we get
\begin{equation}
\begin{split}
q(t) & = \frac{\alpha q_0 + (\omega + \beta) p_0}{R}\sinh(Rt) + q_0\cosh(Rt), \\
p(t) & = -\frac{(\omega - \beta) q_0 + \alpha p_0}{R}\sinh(Rt) + p_0\cosh(Rt),
\end{split}
\end{equation}
and when $\omega^2 = \alpha^2 + \beta^2$
\begin{equation}
\begin{split}
q(t) & = q_0 + (\alpha q_0 + (\omega + \beta) p_0)t, \\
p(t) & = p_0 - ((\omega - \beta) q_0 + \alpha p_0)t.
\end{split}
\end{equation}

First, let us focus on the case $\omega^2 > \alpha^2 + \beta^2$. We receive the
following formulas for time evolution of observables of position and momentum
\begin{equation}
\begin{split}
Q(q,p,t) & = \frac{\alpha q + (\omega + \beta) p}{R}\sin(Rt) + q\cos(Rt), \\
P(q,p,t) & = -\frac{(\omega - \beta) q + \alpha p}{R}\sin(Rt) + p\cos(Rt).
\end{split}
\end{equation}
Again the above formulas describe time evolution of $Q$ and $P$ in classical as
well as in quantum case, which in other words means that classical and quantum
trajectories coincide. Also for this quantum system time evolution of
expectation values of position and momentum observables in coherent states
coincide with quantum trajectories, as is the case in classical theory.

\begin{figure*}
\centering
\begin{tabular}{cc}
\includegraphics{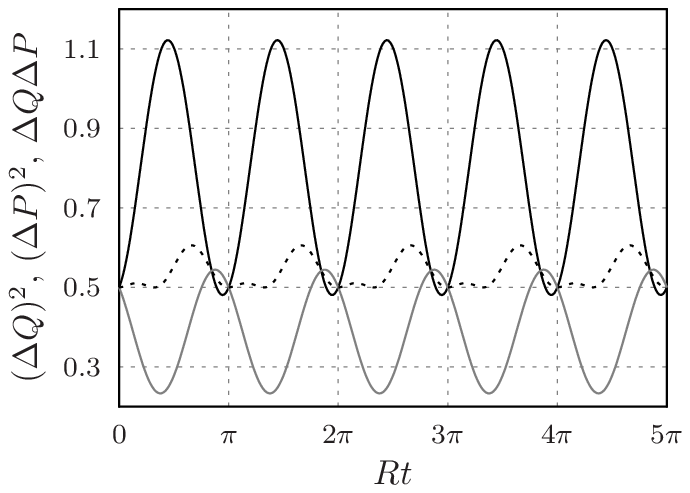} & \includegraphics{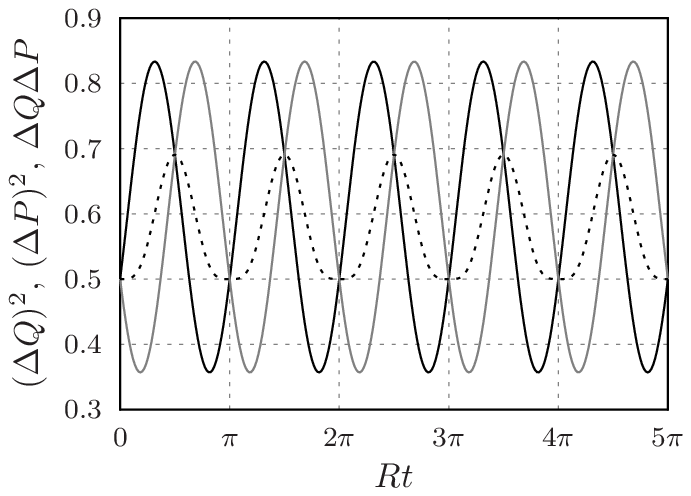} \\
(a) & (b) \\
\includegraphics{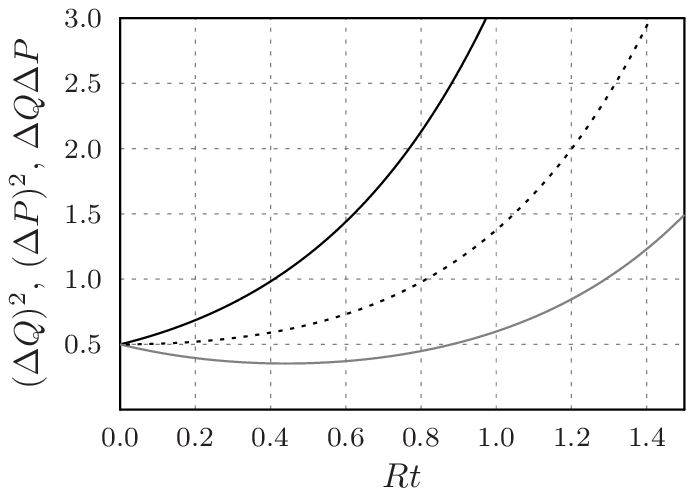} & \includegraphics{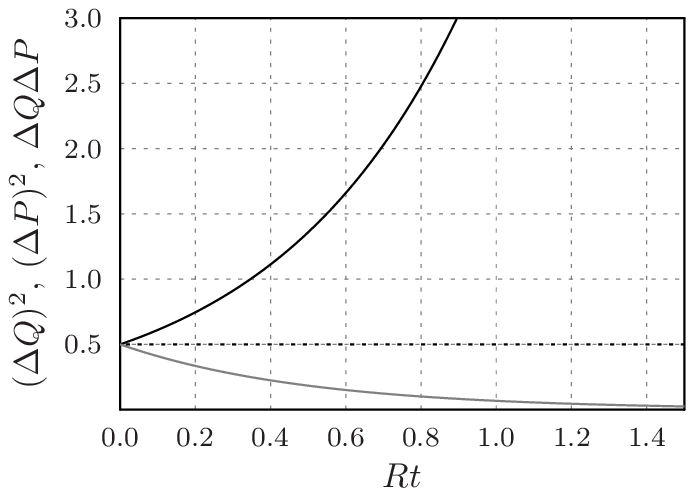} \\
(c) & (d)
\end{tabular}
\caption{Plots of $(\Delta Q)^2$ (black), $(\Delta P)^2$ (gray) and
$\Delta Q \Delta P$ (dotted black) by $Rt$ for $\hbar = \gamma = 1$ and
(a) $r = 2$, $\theta = \frac{\pi}{3}$, $\omega = 5$ (plots of \eqref{eq:12} and
\eqref{eq:13}), (b) $r = 2$, $\theta = 0$, $\omega = 5$ (plots of \eqref{eq:14}
and \eqref{eq:15}), (c) $r = 2$, $\theta = \frac{\pi}{4}$, $\omega = 0$
(plots of \eqref{eq:16} and \eqref{eq:17}), (d) $r = 2$, $\theta = 0$,
$\omega = 0$ (plots of \eqref{eq:18} and \eqref{eq:19}).}
\label{fig:1}
\end{figure*}

In an ideal squeezed state
\begin{equation}
\rho(q,p) = 2\exp\left(-\frac{(q - q_0)^2}{\hbar\gamma^{-1}}\right)
    \exp\left(-\frac{(p - p_0)^2}{\hbar\gamma}\right),
\label{eq:20}
\end{equation}
where $\gamma > 0$ is a parameter describing a squeezing of the state,
we receive the following formulas for the uncertainties $(\Delta Q)^2$ and
$(\Delta P)^2$
\begin{equation}
\begin{split}
(\Delta Q)^2 & = \frac{\hbar}{2}\gamma^{-1}\left(1 + \frac{\sin^2(Rt)}{R}
    \left(\frac{2r^2 + 2\gamma^2\omega r\sin\theta
    + (\gamma^2 - 1)(\omega^2 + r^2 \sin^2\theta)}{R} + 2r\cos\theta \cot(Rt)
    \right) \right), \\
(\Delta P)^2 & = \frac{\hbar}{2}\gamma\left(1 + \frac{\sin^2(Rt)}{R}
    \left(\frac{2r^2 - 2\gamma^{-2}\omega r\sin\theta
    + (\gamma^{-2} - 1)(\omega^2 + r^2 \sin^2\theta)}{R} - 2r\cos\theta \cot(Rt)
    \right) \right),
\end{split}
\end{equation}
where we introduced new parameters $r,\theta$ such that $\alpha + i\beta =
re^{i\theta}$. From this we can see that for certain values of the evolution
parameter $t$ the ideal squeezed state \eqref{eq:20} evolves into a squeezed
state. It can be calculated that in the case $\gamma = 1$
\begin{equation}
\begin{split}
(\Delta Q)^2 & = \frac{\hbar}{2}\left(1 + \frac{2r}{R} \sin^2(Rt)
    \left(\frac{r + \omega \sin\theta}{R} + \cos\theta \cot(Rt)
    \right) \right), \\
(\Delta P)^2 & = \frac{\hbar}{2}\left(1 + \frac{2r}{R} \sin^2(Rt)
    \left(\frac{r - \omega \sin\theta}{R} - \cos\theta \cot(Rt)
    \right) \right),
\end{split}
\label{eq:12}
\end{equation}
and
\begin{equation}
(\Delta Q)^2 (\Delta P)^2 = \frac{\hbar^2}{4}\left(1 + \frac{4r^2}{R^2}
    \sin^2(Rt) \left(\frac{\omega}{R}\cos\theta \sin(Rt) - \sin\theta \cos(Rt)
    \right)^2 \right).
\label{eq:13}
\end{equation}
From \eqref{eq:13} we can see that the minimization of the Heisenberg
uncertainty relation occurs only for $Rt = k\pi$ and
$Rt = \arctan(\frac{R}{\omega}\tan\theta) + k\pi$, $k \in \mathbb{Z}$.
In fig.~\ref{fig:1} are presented plots of $(\Delta Q)^2$, $(\Delta P)^2$ and
$\Delta Q \Delta P$ for certain values of parameters $\omega$, $r$, $\theta$.
From plots (a) and (b) we can see that there are intervals of $t$ on which
$(\Delta Q)^2 < \frac{\hbar}{2}$ or $(\Delta P)^2 < \frac{\hbar}{2}$ but there
are also intervals on which the state is not squeezed.

In what follows we will consider several particular cases of the choice of
parameters $\omega,\alpha,\beta,\gamma$. First let us take $\beta = 0$ and
$\alpha > 0$. Then $r = \alpha$ and $\theta = 0$. In this case we get that
\begin{equation}
\begin{split}
(\Delta Q)^2 & = \frac{\hbar}{2}\gamma^{-1}\left(1 + \frac{\sin^2(Rt)}{R}
    \left(\frac{2r^2 + (\gamma^2 - 1)\omega^2}{R} + 2r\cot(Rt)\right)\right), \\
(\Delta P)^2 & = \frac{\hbar}{2}\gamma\left(1 + \frac{\sin^2(Rt)}{R}
    \left(\frac{2r^2 + (\gamma^{-2} - 1)\omega^2}{R} - 2r\cot(Rt)\right)\right),
\end{split}
\end{equation}
and
\begin{align}
(\Delta Q)^2 (\Delta P)^2 = \frac{\hbar^2}{4}\biggl( &
    1 + (\gamma^2 + \gamma^{-2}) \frac{2\omega^2r^2}{R^4} \sin^4(Rt)
    + (\gamma^2 + \gamma^{-2} - 2) \frac{\omega^2}{R^2}\sin^2(Rt) \nonumber \\
& {} - (\gamma^2 - \gamma^{-2}) \frac{2\omega^2 r}{R^3} \sin^3(Rt)\cos(Rt)
    \biggr).
\end{align}
If additionally $\gamma = 1$ we receive
\begin{equation}
\begin{split}
(\Delta Q)^2 & = \frac{\hbar}{2}\left(1 + \frac{2r}{R}\sin^2(Rt)
    \left(\frac{r}{R} + \cot(Rt)\right)\right), \\
(\Delta P)^2 & = \frac{\hbar}{2}\left(1 + \frac{2r}{R}\sin^2(Rt)
    \left(\frac{r}{R} - \cot(Rt)\right)\right),
\end{split}
\label{eq:14}
\end{equation}
and
\begin{equation}
(\Delta Q)^2 (\Delta P)^2 = \frac{\hbar^2}{4}\left(1 + \frac{4\omega^2 r^2}{R^4}
    \sin^4(Rt) \right).
\label{eq:15}
\end{equation}

Now, let us consider the case $\omega^2 < \alpha^2 + \beta^2$ taking
$\omega = 0$. Then $R = r$ and
\begin{equation}
\begin{split}
Q(q,p,t) & = (q\cos\theta + p\sin\theta)\sinh(rt) + q\cosh(rt), \\
P(q,p,t) & = (q\sin\theta - p\cos\theta)\sinh(rt) + p\cosh(rt).
\end{split}
\end{equation}
Moreover,
\begin{equation}
\begin{split}
(\Delta Q)^2 & = \frac{\hbar}{2}\gamma^{-1}\bigl(\cosh(2rt)
    + \cos\theta \sinh(2rt) + (\gamma^2 - 1)\sin^2\theta \sinh^2(rt)\bigr), \\
(\Delta P)^2 & = \frac{\hbar}{2}\gamma\bigl(\cosh(2rt) - \cos\theta \sinh(2rt)
    + (\gamma^{-2} - 1)\sin^2\theta \sinh^2(rt)\bigr).
\end{split}
\end{equation}
and
\begin{align}
(\Delta Q)^2 (\Delta P)^2 = \frac{\hbar^2}{4}\Bigl( &
    1 + \sin^2\theta \sinh^2(2rt) + (\gamma^2 - 1)\sin^2\theta\sinh^2(rt)
    \bigl(\cosh(2rt) - \cos\theta \sinh(2rt)\bigr) \nonumber \\
& {} + (\gamma^{-2} - 1)\sin^2\theta\sinh^2(rt) \bigl(\cosh(2rt) + \cos\theta
    \sinh(2rt)\bigr) \Bigr).
\end{align}
Again, if additionally $\gamma = 1$ we get
\begin{equation}
\begin{split}
(\Delta Q)^2 & = \frac{\hbar}{2}\bigl(\cosh(2rt) + \cos\theta \sinh(2rt)\bigr)
= \frac{\hbar}{2}\left(e^{2rt} \cos^2\frac{\theta}{2}
    + e^{-2rt} \sin^2\frac{\theta}{2}\right), \\
(\Delta P)^2 & = \frac{\hbar}{2}\bigl(\cosh(2rt) - \cos\theta \sinh(2rt)\bigr)
= \frac{\hbar}{2}\left(e^{-2rt} \cos^2\frac{\theta}{2}
    + e^{2rt} \sin^2\frac{\theta}{2}\right),
\end{split}
\label{eq:16}
\end{equation}
and
\begin{equation}
(\Delta Q)^2 (\Delta P)^2 = \frac{\hbar^2}{4}\left(1 + \sin^2\theta \sinh^2(2rt)
    \right).
\label{eq:17}
\end{equation}
From plot (c) in fig.~\ref{fig:1} we can see that for $\theta \neq k\pi$,
$k \in \mathbb{Z}$ the state is squeezed only at the beginning of time evolution
until a certain value of $t$ is reached. Moreover, as is evident from
\eqref{eq:17} the Heisenberg uncertainty relation is not minimized during the
whole time evolution.

Further on, let us take $\omega = 0$, $\beta = 0$ and $\alpha > 0$. Then
\begin{equation}
\begin{split}
Q(q,p,t) & = q e^{rt}, \\
P(q,p,t) & = p e^{-rt}.
\end{split}
\end{equation}
Moreover,
\begin{equation}
\begin{split}
(\Delta Q)^2 & = \frac{\hbar}{2} \gamma^{-1} e^{2rt}, \\
(\Delta P)^2 & = \frac{\hbar}{2} \gamma e^{-2rt},
\end{split}
\label{eq:18}
\end{equation}
and
\begin{equation}
(\Delta Q)^2 (\Delta P)^2 = \frac{\hbar^2}{4}.
\label{eq:19}
\end{equation}
In this case, if $\gamma \le 1$, the state remains an ideal squeezed state
during the whole time evolution.

Finally, let us consider the case $\omega^2 = \alpha^2 + \beta^2$. In this case
\begin{equation}
\begin{split}
(\Delta Q)^2 & = \frac{\hbar}{2}\left((1 + \alpha t)^2 \gamma^{-1}
    + (\beta + \omega)^2 t^2 \gamma\right), \\
(\Delta P)^2 & = \frac{\hbar}{2}\left((\beta - \omega)^2 t^2 \gamma^{-1}
    + (1 - \alpha t)^2 \gamma\right),
\end{split}
\end{equation}
and
\begin{equation}
(\Delta Q)^2 (\Delta P)^2 = \frac{\hbar^2}{4}\bigl(1 - 2\alpha^2 t^2
    + 2\alpha^4 t^4 + (1 + \alpha t)^2 (\beta - \omega)^2 t^2 \gamma^{-2}
    + (1 - \alpha t)^2 (\beta + \omega)^2 t^2 \gamma^2\bigr).
\end{equation}
It can be seen that during time development the squeezing decreases in time
until it is completely destroyed at certain instance of time. Moreover, the
Heisenberg uncertainty relation is not minimized during the whole time
evolution.

Note, that after performing the following transformation to new coordinates
(rotation by angle $\frac{\theta}{2}$)
\begin{equation}
\begin{split}
q' & = q\cos\frac{\theta}{2} + p\sin\frac{\theta}{2}, \\
p' & = -q\sin\frac{\theta}{2} + p\cos\frac{\theta}{2}
\end{split}
\end{equation}
the Hamiltonian $H(q,p)$ \eqref{eq:24} transforms into
\begin{equation}
H(q',p') = \frac{1}{2}\omega(p'^2 + q'^2) + rq'p'.
\end{equation}
Thus, we can always perform the reduction to the case $\beta = 0$ and
$\alpha > 0$ provided that we will be working with rotated variables $q',p'$.
Furthermore, in a case $\omega^2 > \alpha^2 + \beta^2$ the following
one-parameter family of linear canonical transformations of coordinates
\begin{equation}
\begin{split}
q' & = \frac{Ra + \alpha A}{\omega + \beta}q + Ap, \\
p' & = \frac{\alpha a - RA}{\omega + \beta}q + ap,
\end{split}
\quad A = \pm\sqrt{\frac{\omega + \beta}{R} - a^2}, \quad a \in \mathbb{R}
\end{equation}
transforms the Hamiltonian \eqref{eq:24} into the following Hamiltonian of the
harmonic oscillator with positive frequency
\begin{equation}
H(q',p') = \frac{1}{2}R(p'^2 + q'^2),
\end{equation}
and the one-parameter family of transformations
\begin{equation}
\begin{split}
q' & = -\frac{Ra + \alpha A}{\omega + \beta}q - Ap, \\
p' & = \frac{\alpha a - RA}{\omega + \beta}q + ap,
\end{split}
\quad A = \pm\sqrt{-\frac{\omega + \beta}{R} - a^2}, \quad a \in \mathbb{R}
\end{equation}
leads to the following Hamiltonian of the harmonic oscillator with negative
frequency
\begin{equation}
H(q',p') = -\frac{1}{2}R(p'^2 + q'^2).
\end{equation}
On the other hand, in a case $\omega^2 < \alpha^2 + \beta^2$ the following
one-parameter family of linear canonical transformations of coordinates
\begin{equation}
\begin{split}
q' & = -\frac{R + \alpha}{\omega + \beta}aq - ap, \\
p' & = \frac{R - \alpha}{2Ra}q - \frac{\omega + \beta}{2Ra}p,
\end{split}
\quad a \in \mathbb{R}
\end{equation}
transforms the Hamiltonian \eqref{eq:24} into the following Hamiltonian
\begin{equation}
H(q',p') = Rq'p',
\end{equation}
and the one-parameter family of transformations
\begin{equation}
\begin{split}
q' & = \frac{R - \alpha}{\omega + \beta}aq - ap, \\
p' & = \frac{R + \alpha}{2Ra}q + \frac{\omega + \beta}{2Ra}p,
\end{split}
\quad a \in \mathbb{R}
\end{equation}
leads to the following Hamiltonian
\begin{equation}
H(q',p') = -Rq'p'.
\end{equation}

Summarizing the above considerations we conclude that when
$\omega^2 > \alpha^2 + \beta^2$ we can always reduce the dynamics to the
harmonic oscillator, and when $\omega^2 < \alpha^2 + \beta^2$ we can reduce the
dynamics to the case $\omega = \beta = 0$, provided that we will be working with
new variables $q',p'$. In the frame of original variables $q,p$ it means that
for the considered system there always exist canonically conjugated observables
$q' = q'(q,p), p' = p'(q,p)$ which preserve the minimal uncertainty during time
evolution.

\section{Hamiltonian system with purely quantum time evolution}
\label{sec:4}
Let us consider a system described by a Hamiltonian
\begin{equation}
H(q,p) = \lambda q^2 p^2,
\label{eq:1}
\end{equation}
where $\lambda$ is a characteristic constant of the system. By a coherent state
of such system we will call a state for which uncertainties of position and
momentum satisfy \eqref{eq:10}. Coherent states are of the form of Gaussian
functions
\begin{equation}
\rho(q,p) = 2\exp\left(-\frac{(q - q_0)^2}{\hbar\alpha}\right)
    \exp\left(-\frac{(p - p_0)^2}{\hbar\alpha^{-1}}\right),
\label{eq:3}
\end{equation}
where $(q_0,p_0)$ is a mean position and momentum in the state $\rho$, and are
parametrized by $\alpha > 0$.

The solution of the quantum Hamilton equations \eqref{eq:6} for the Hamiltonian
\eqref{eq:1} reads \cite{Dias:2007}
\begin{equation}
\begin{split}
Q(q,p,t) & = \sec^2(\hbar \lambda t) q \exp\left(\frac{2}{\hbar}
    \tan(\hbar \lambda t) qp\right), \\
P(q,p,t) & = \sec^2(\hbar \lambda t) p \exp\left(-\frac{2}{\hbar}
    \tan(\hbar \lambda t) qp\right).
\end{split}
\label{eq:2}
\end{equation}
It is well defined for $t \in \mathbb{R} \setminus \{\frac{\pi n}{2\hbar\lambda}
\mid n = \pm 1,\pm 2,\dotsc\}$. Note, that this solution differs from the
classical one
\begin{equation}
Q_C(q,p,t) = q e^{2\lambda tqp}, \quad P_C(q,p,t) = p e^{-2\lambda tqp},
\end{equation}
which can be obtained from \eqref{eq:2} in the limit $\hbar \to 0$. In other
words quantum trajectories differ from classical ones. Moreover, time
development of position and momentum observables is not well defined for
all values of the evolution parameter $t$, contrary to the classical case.

Through direct integration we can calculate the expectation values of $Q$ and
$P$ from \eqref{eq:2} in a coherent state \eqref{eq:3}. The result after
introducing
\begin{equation}
\begin{split}
a(t) & = \frac{\cos(\hbar\lambda t)}{\sqrt{\cos(2\hbar\lambda t)}} q_0
    + \frac{\alpha \sin(\hbar\lambda t)}{\sqrt{\cos(2\hbar\lambda t)}} p_0, \\
b(t) & = -\frac{\alpha^{-1} \sin(\hbar\lambda t)}{\sqrt{\cos(2\hbar\lambda t)}}
    q_0 + \frac{\cos(\hbar\lambda t)}{\sqrt{\cos(2\hbar\lambda t)}} p_0,
\end{split}
\end{equation}
reads
\begin{equation}
\begin{split}
\braket{Q}_\rho & = \frac{a(t)}{\cos(2\hbar\lambda t)}
    \exp\biggl(\frac{1}{\hbar\alpha}\left(a^2(t) - q_0^2\right)\biggr), \\
\braket{P}_\rho & = \frac{b(t)}{\cos(2\hbar\lambda t)}
    \exp\biggl(\frac{\alpha}{\hbar}\left(b^2(t) - p_0^2\right)\biggr).
\end{split}
\end{equation}
Note, that $\braket{Q}_\rho$ and $\braket{P}_\rho$ are well defined only on
intervals $(-\frac{1}{4} + n)\frac{\pi}{\hbar\lambda} < t < (\frac{1}{4} + n)
\frac{\pi}{\hbar\lambda}$, $n \in \mathbb{Z}$. This once again shows that
time evolution of the considered system is not defined for all values of the
evolution parameter $t$ and even time development of quantities measured in
experiment like expectation values of position and momentum is only well defined
on certain intervals of $t$. Observe also, that time evolution of expectation
values of $Q$ and $P$ in a coherent state does not coincide with quantum
trajectories, as was the case for linear quantum systems.

Let us calculate standard deviations of $Q$ and $P$ in the coherent state
$\rho$. To do this first we need to calculate expectation values of
$Q^2 = Q \star_M Q$ and $P^2 = P \star_M P$. To calculate $Q^2$ we can perform
Wick rotation $t \to i\tau$ and use the integral formula \eqref{eq:5} for the
Moyal product. We find that
\begin{equation}
(Q \star_M Q)(q,p,t) = \sec^3(2\hbar\lambda t) q^2 \exp\left(\frac{2}{\hbar}
    \tan(2\hbar\lambda t) qp\right).
\end{equation}
Similarly we calculate $P^2$
\begin{equation}
(P \star_M P)(q,p,t) = \sec^3(2\hbar\lambda t) p^2 \exp\left(-\frac{2}{\hbar}
    \tan(2\hbar\lambda t) qp\right).
\end{equation}

The expectation values of $Q^2$ and $P^2$ can be calculated similarly as
$\braket{Q}_\rho$ and $\braket{P}_\rho$. We get the following formulas
\begin{equation}
\begin{split}
\braket{Q^2}_\rho & = \bigl(\cos(4\hbar\lambda t)\bigr)^{-3/2}
    \left(\frac{\hbar}{2}\alpha + a^2(2t)\right)
    \exp\biggl(\frac{1}{\hbar\alpha} \left(a^2(2t) - q_0^2\right)
    \biggr), \\
\braket{P^2}_\rho & = \bigl(\cos(4\hbar\lambda t)\bigr)^{-3/2}
    \left(\frac{\hbar}{2}\alpha^{-1} + b^2(2t)\right)
    \exp\biggl(\frac{\alpha}{\hbar} \left(b^2(2t) - p_0^2\right)
    \biggr),
\end{split}
\end{equation}
which are well defined for $(-\frac{1}{4} + n)\frac{\pi}{2\hbar\lambda} < t <
(\frac{1}{4} + n)\frac{\pi}{2\hbar\lambda}$, $n \in \mathbb{Z}$.

The standard deviations of $Q$ and $P$ are equal
\begin{equation}
\begin{split}
(\Delta Q)^2 & = \bigl(\cos(4\hbar\lambda t)\bigr)^{-3/2}
    \left(\frac{\hbar}{2}\alpha + a^2(2t)\right)
    \exp\biggl(\frac{1}{\hbar\alpha}\left(a^2(2t) - q_0^2\right)\biggr)
    - \frac{a^2(t)}{\cos^2(2\hbar\lambda t)}
    \exp\biggl(\frac{2}{\hbar\alpha}\left(a^2(t) - q_0^2\right)\biggr), \\
(\Delta P)^2 & = \bigl(\cos(4\hbar\lambda t)\bigr)^{-3/2}
    \left(\frac{\hbar}{2}\alpha^{-1} + b^2(2t)\right)
    \exp\biggl(\frac{\alpha}{\hbar}\left(b^2(2t) - p_0^2\right)\biggr)
    - \frac{b^2(t)}{\cos^2(2\hbar\lambda t)}
    \exp\biggl(\frac{2\alpha}{\hbar}\left(b^2(t) - p_0^2\right)\biggr)
\end{split}
\label{eq:7}
\end{equation}
and are well defined for $(-\frac{1}{8} + n)\frac{\pi}{\hbar\lambda} < t <
(\frac{1}{8} + n)\frac{\pi}{\hbar\lambda}$, $n \in \mathbb{Z}$.

\begin{figure}
\centering
\includegraphics{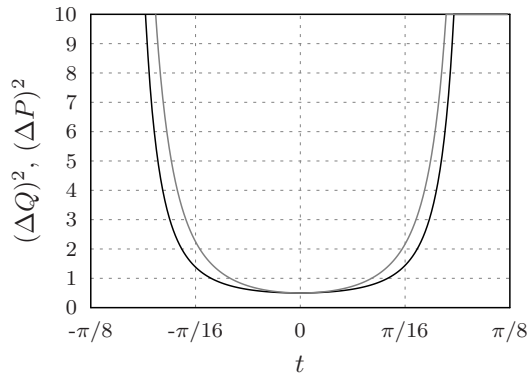}
\caption{Plots of $(\Delta Q)^2$ (black) and $(\Delta P)^2$ (gray) by $t$ given
by equation \eqref{eq:7} for $\hbar = \lambda = \alpha = 1$, $q_0 = 0.01$,
$p_0 = 1$.}
\label{fig:2}
\end{figure}

Note, that there are no values of the parameters $q_0,p_0,\alpha$ such that
$(\Delta Q)^2 (\Delta P)^2$ is equal $\frac{\hbar^2}{4}$ for
all $t$. In other words the coherent state $\rho$ does not remain coherent
during time development. It happens to be coherent only for
$t = \frac{\pi n}{\hbar\lambda}$, $n \in \mathbb{Z}$. Moreover, the state $\rho$
is not getting squeezed during time evolution in the sense that $(\Delta Q)^2$
or $(\Delta P)^2$ is not getting smaller than its initial value. In
fig.~\ref{fig:2} are presented plots of $(\Delta Q)^2$ and $(\Delta P)^2$
for certain values of parameters $\lambda,\alpha,q_0,p_0$.

\section{Conclusions and final remarks}
\label{sec:5}
In the paper we have investigated the behavior of coherent and squeezed states
during time development generated by two distinct Hamiltonians. As the first
Hamiltonian we took a Hamiltonian quadratic in phase space variables $q,p$.
This was one of the simplest Hamiltonians for which classical and quantum flows
coincided, and even in this case the initial squeezed state remained squeezed
only on certain intervals of $t$, and only for certain values of $t$ the state
was coherent. In fact, only for particular choice of parameters we get the
preservation of coherence and squeezing during the whole time evolution. This
result allows to suspect that for more general Hamiltonians the coherence and
squeezing of states will rarely appear during time development. Furthermore,
we showed that after performing appropriate linear canonical transformation of
phase space coordinates we could always transform the Hamiltonian quadratic in
phase space variables to a Hamiltonian for which the coherence and squeezing
will be preserved during time development, provided that the coherence and
squeezing will be defined with respect to the transformed phase space
coordinates.

As the second example we considered the simplest Hamiltonian with purely quantum
time evolution. In this case the coherence was destroyed during time development
and squeezing did not appear. Moreover, the quantum trajectories and expectation
values of observables of position and momentum were well defined only on certain
intervals of $t$, which raises problems and questions of interpretation of such
kind of time evolution. This could suggest that the Moyal quantization (Weyl
quantization in position representation) applied for the system \eqref{eq:1}
is not the proper quantization for such system, and that one should use some
other quantization in which at least expectation values of observables of
position and momentum will be well defined for all values of $t$. Another
question is whether in considered quantization there are some quantum states in
which these expectation values are properly defined for all $t$ (states
\eqref{eq:3} does not have such property). We can suspect that for more general
quantum systems even more peculiarities can appear, which is an interesting
topic for further investigation.

\section*{Acknowledgments}
Z.~Doma\'nski acknowledge the support of Polish National Science Center grant
under the contract number DEC-2011/02/A/ST1/00208.


%

\end{document}